\documentstyle[12pt]{article}

\newcommand{\be}{\begin{equation}}
\newcommand{\ee}{\end{equation}}
\newcommand{\bi}[1]{\vspace{-3mm} \bibitem{#1}}

\begin{document}

\begin{center}
{\it Journal of Physics A 37 (2004) 3241-3257}

\vspace{7mm}

{\Large \bf Path Integral for Quantum Operations} \\
\vspace{4mm}
{\large \bf Vasily E. Tarasov}\\
Skobeltsyn Institute of Nuclear Physics,
Moscow State University, 119992 Moscow, Russia \\
E-mail: tarasov@theory.sinp.msu.ru
\end{center}

\begin{abstract}
In this paper we consider a phase space path integral
for general time-dependent quantum operations,
not necessarily unitary.
We obtain the path integral for a completely
positive quantum operation satisfied Lindblad equation
(quantum Markovian master equation).
We consider the path integral for quantum operation with
a simple infinitesimal generator.
\end{abstract}

PACS 03.67.Lx, 03.067-a, 03.65.-w




\section{Introduction}


Unitary evolution is not the most general type
of state change possible for quantum systems.
The most general state change of a quantum system
is a quantum operation \cite{Kr1,Kr2,Kr3,Kr4,Schu}.
One can describe a quantum operation for a quantum system starting
from a unitary evolution of some closed system if
the quantum system is a part of the closed system \cite{FV}-\cite{Str}.
However, situations can arise where it is difficult or impossible 
to find a closed system comprising the given quantum system \cite{T2}-\cite{kn1}.
This would render theory of quantum operations
a fundamental generalization of the unitary evolution
of the closed quantum system.

The usual models of a quantum computer deal only with unitary
quantum operations on pure states.
In these models it is difficult or impossible
to deal formally with measurements, dissipation, decoherence and noise.
It turns out that the restriction to pure states and unitary gates
is unnecessary \cite{TarJPA}.
In \cite{TarJPA}, a model of quantum computations
by quantum operations with mixed states was constructed.
The computations are realized by quantum operations,
not necessarily unitary. 
Mixed states subjected to general quantum operations could increase efficiency.
This increase is connected with the increasing number of
computational basis elements for the Hilbert space.
A pure state of n two level quantum systems is an element
of the $2^n$-dimensional functional Hilbert space.
A mixed state of the system is an element of
the $4^n$-dimensional operator Hilbert space.
Therefore, the increased efficiency can be formalized in terms of a
four-valued logic replacing the conventional two-valued logic.
Unitary gates and quantum operations for a quantum computer with pure states
can be considered as quantum gates of a mixed state quantum computer.
Quantum algorithms on a quantum computer with mixed states are expected
to run on a smaller network than with pure state implementation.

The path integral for quantum operations can be useful for the
continuous-variable generalization of quantum computations
by quantum operations with mixed states.
The usual models of a quantum computer deal only with
discrete variables.
Many quantum variables such as position and momentum are continuous.
The use of continuous-variable quantum computing \cite{CV1,CV2,CV3}
allows information to be encoded and processed much more compactly
and efficiently than with discrete variable computing.
Quantum computation using continuous variables is an alternative
approach to quantum computations with discrete variables.

All processes occur in time. 
It is naturally to consider time dependence for quantum operations.
In this paper we consider the path integral approach to
general time-dependent quantum operations.
We use the operator space \cite{Cra}-\cite{kn2} and
superoperators on this space.
The path integral for unitary evolution from the operator (Liouville)
space was derived in \cite{Schm2}.
The quantum operation is considered as a real completely positive
trace-preserving superoperator on the operator space.
We derive a path integral for a completely positive
quantum operation satisfied Lindblad equation
(quantum Markovian master equation) \cite{Lind,K1,K2,AL,T6,kn2}.
For example, we consider a path integral for a quantum operation
with a simple infinitesimal generator.

In section 2, the requirements for a superoperator to be
a generalized quantum operation are discussed.
In section 3, the general Liouville-von Neumann equation and
quantum Markovian (Lindblad) master equation are considered.
In section 4, we derive path integral
for quantum operation satisfied
Liouville-von Neumann equation.
In section 5, we obtain a path integral for
time-dependent quantum operation with an infinitesimal generator
such that the adjoint generator is completely dissipative.
In section 6, the continuous-variables quantum computation 
by quantum operations with mixed states is discussed.
In the appendix, the mathematical  background
(Liouville space, superoperators) is considered.


\section{Quantum operations as superoperators}

Unitary evolution is not the most general type
of state change possible for quantum systems.
The most general state change of a quantum system
is a positive trace-preserving map  which
is called a quantum operation.
For the concept of quantum operations,
see \cite{Kr1,Kr2,Kr3,Kr4,Schu}.

A quantum operation is a superoperator $\hat{\cal E}$
which maps the density matrix operator $|\rho)$ 
to the density matrix operator $|\rho^{\prime})$.
For the concept of superoperators and operator space see
the appendix and \cite{Cra}-\cite{kn2}

If $|\rho)$ is a density matrix operator, then $\hat{\cal E}|\rho)$
should also be a density matrix operator.
Any density matrix operator $\rho$ is a
self-adjoint ($\rho^{\dagger}_{t}=\rho_{t}$),
positive ($\rho_{t}>0$) operator with unit trace ($ Tr\rho_{t}=1$).
Therefore, the requirements for a superoperator $\hat{\cal E}$
to be the quantum operation are as follows:
\begin{enumerate}
\item
The superoperator $\hat{\cal E}$ is a {\it real} superoperator, i.e.
$\Bigl(\hat{\cal E}(A)\Bigr)^{\dagger}=\hat{\cal E}(A^{\dagger})$
for all $A$.
The real superoperator $\hat{\cal E}$ maps the self-adjoint operator
$\rho$ to the self-adjoint operator $\hat{\cal E}(\rho)$:
\ $(\hat{\cal E}(\rho))^{\dagger}=\hat{\cal E}(\rho)$.
\item
The superoperator $\hat{\cal E}$ is a {\it positive} superoperator,
i.e. $\hat{\cal E}$ maps positive operators to positive operators:
\ $\hat{\cal E}(A^{2}) >0$ for all $A\not=0$ or
$\hat{\cal E}(\rho)\ge 0$.
\item
The superoperator $\hat{\cal E}$ is a {\it trace-preserving} map, i.e.
$(I|\hat{\cal E}|\rho)=(\hat{\cal E}^{\dagger}(I)|\rho)=1$
or $\hat{\cal E}^{\dagger}(I)=I$.
\end{enumerate}

We have to assume the superoperator $\hat{\cal E}$
to be not merely positive but completely positive \cite{Arveson}.
The superoperator $\hat{\cal E}$ is a {\it completely positive}
map of the operator space, if
\[ \sum^{n}_{k=1} \sum^{n}_{l=1} B^{\dagger}_{k}
\hat{\cal E}(A^{\dagger}_kA_l)B_l \ge 0 \]
for all operators $A_k$, $B_k$ and all $n$.\\
Let the superoperator $\hat{\cal E}$ be a {\it convex linear} map
on the set of density matrix operators, i.e.
\[ \hat{\cal E}\Bigl(\sum_{s} \lambda_{s} \rho_{s}\Bigr)=
\sum_{s} \lambda_{s} \hat{\cal E}(\rho_{s}), \]
where all $\lambda_{s}$ are $0<\lambda_{s}<1$ and $\sum_{s} \lambda_{s}=1$.
Any convex linear map of density matrix operators
can be uniquely extended to a {\it linear} map on Hermitian operators.
Note that any linear completely positive superoperator can be represented by
\[ \hat{\cal E}=\sum^{m}_{k=1} \hat L_{A_{k}} \hat R_{A^{\dagger}_{k}}: \quad
\hat{\cal E}(\rho)=\sum^{m}_{k=1} A_k \rho A^{\dagger}_k. \]
If this superoperator is a trace-preserving superoperator, then
\[ \sum^{m}_{k=1} A^{\dagger}_{k} A_{k}=I. \]

The restriction to linear quantum operations is unnecessary.
Let us consider a linear real completely positive
superoperator $\hat{\cal E}$ which is not trace-preserving.
Let $(I|\hat{\cal E}|\rho)=Tr(\hat{\cal E}(\rho))$
be the probability that the process represented
by the superoperator $\hat{\cal E}$ occurs.
Since the probability is non-negative and
never exceed 1, it follows that the superoperator
$\hat{\cal E}$ is a trace-decreasing superoperator:
$0\le (I|\hat{\cal E}|\rho) \le 1$ or $\hat{\cal E}^{\dagger}(I) \le I$.
In general, any real linear completely positive trace-decreasing
superoperator is not a quantum operation, since
it can be not trace-preserving.
The quantum operation cannot be defined as a {\it nonlinear trace-preserving} 
operation $\hat{\cal N}$ by
\be \label{N} \hat{\cal N}|\rho)=\hat{\cal E}|\rho)
(I|\hat{\cal E}|\rho)^{-1}\quad or
\quad \hat{\cal N}(\rho)=
\frac{\hat{\cal E}(\rho)}{Tr(\hat{\cal E}(\rho))}, \ee
where $\hat{\cal E}$ is a real linear completely positive
trace-decreasing superoperator.

All processes occur in time. It is naturally to consider time
dependence for quantum operations $\hat {\cal E}(t,t_{0})$. Let the
linear superoperators $\hat {\cal E}(t,t_{0})$ form a completely
positive quantum semigroup \cite{AL} such that \be \label{qo1}
\frac{d}{dt}\hat{\cal E}(t,t_{0})= \hat\Lambda_t \hat{\cal
E}(t,t_{0}), \ee
where $\hat \Lambda^{\dagger}$ is a completely
dissipative superoperator \cite{Lind,AL,kn1}. 
We would like to consider the path integral for quantum operations 
$\hat{\cal E}(t,t_0)$ with infinitesimal generator $\hat \Lambda$, where
the adjoint superoperator $\hat \Lambda^{\dagger}$ is 
{\it completely dissipative}, i.e.
\[\hat \Lambda^{\dagger}(A_{k}A_{l})-
\hat \Lambda^{\dagger}(A_{k}) A_{l}-
A_{k}\hat \Lambda^{\dagger}(A_{l}) \ge 0. \]


\section{Evolution equations}


An important property of most open and dissipative quantum systems
is the entropy variation. Nevertheless the unitary quantum evolution
of a mixed state $\varrho_t$ described by von Neumann equation
\be \label{neum}
\frac{\partial \varrho_{t}}{\partial t}
=-\frac{i}{\hbar}[H,\varrho_{t}] \ee
leaves the entropy $<S>=-Tr(\varrho_t \ ln\varrho_t)$ unchanged.
Therefore, to describe general quantum systems, one normally
uses \cite{Dek,kn2} a generalization of (\ref{neum}).

To describe dissipative quantum systems one usually considers \cite{Dek}
the following equation: 
\be \label{neum3} \frac{\partial \varrho_{t}}{\partial t}
=-\frac{i}{\hbar}[H,\varrho_{t}]+D(\varrho) . \ee


\subsection{Liouville-von Neumann equation}



Let us consider a generalization of equation (\ref{neum}).
The Liouville-von Neumann equation \cite{T3,kn2,T7} can be
represented as the linear equation
\be \label{v-r-1}
\frac{d \varrho_t}{dt}=\Lambda_{t} (\varrho_t) . \ee
Using the superoperator formalism, this equation can be rewritten in the form
\be \label{v-r-2}
\frac{d}{dt} |\varrho_t)=\hat \Lambda_{t} |\varrho_t) . \ee
The superoperator language allows one to use the analogy with Dirac's notations.
This leads quite simple to the derivation of the appropriate equations.

Here $\hat \Lambda_{t}$ is a linear Liouville superoperator
on the operator space $\overline{\cal H}$.
For the Hamiltonian (closed) quantum systems (\ref{neum})
this superoperator is defined by the Hamiltonian $H$:
\be \hat \Lambda_{t}
=-\frac{i}{\hbar}(\hat L_H-\hat R_H) . \ee
For equation (\ref{neum3}) the Liouville superoperator has the form
\be \hat \Lambda_{t}
=-\frac{i}{\hbar}(\hat L_H-\hat R_H)+\hat D . \ee
In general, the operator $| \varrho_{t})$ is
an unnormalized density matrix operator, i.e.
$Tr \varrho_{t}=(I|\varrho_{t})\not=1$.

Equation (\ref{v-r-2}) has a formal solution
\be \label{s181}
|\varrho_t)=\hat{\cal E}(t,t_0)|\varrho_{t_0}), \ee
where $\hat{\cal E}(t,t_0)$ is a linear quantum operation defined by
\be \label{qo2} \hat{\cal E}(t,t_0)=Texp\int^{t}_{t_0} d\tau \hat
\Lambda_{\tau}. \ee The symbol $T$ is a Dyson's time-ordering
operator \cite{Dys1}. The quantum operation (\ref{qo2}) satisfies
the Liouville-von Neumann equation (\ref{qo1}).
We can define a normalized density matrix operator $|\rho_{t})$ by
\[ |\rho_t)=|\varrho_t) (I|\varrho_t)^{-1}, \ \ or
\ \ |\rho_t)=\frac{\hat{\cal E}(t,t_{0})|\varrho)}{
(I|\hat{\cal E}(t,t_{0})|\varrho)}, \]
i.e. $\rho_t=\varrho_t / Tr \varrho_t$.
The evolution equation for the normalized density
matrix operator $\rho_{t}$ can be written in the form
\be \label{nle} \frac{d}{dt}|\rho_t)=
\hat \Lambda_{t}|\rho_t)-|\rho_t)(I|\hat \Lambda_{t}|\rho_t). \ee
In general, this equation is a nonlinear equation \cite{kn1}.
A formal solution of equation (\ref{nle}) is connected
with the nonlinear quantum operation (\ref{N}) by
\[ |\rho_{t})=\hat{\cal N}(t,t_0)|\rho_{t_0}). \]



\subsection{Quantum Markovian equation}

Lindblad \cite{Lind} has shown that there exists a one-to-one
correspondence between the completely positive norm continuous
semigroup of superoperators $\hat{\cal E}(t,t_0)$ and
superoperator $\hat \Lambda$ such that the adjoint superoperator
$\hat \Lambda^{\dagger}$ is completely dissipative.
The structural theorem of Lindblad gives the most general form of the bounded
adjoint completely dissipative Liouville superoperator $\hat \Lambda$.
The Liouville-von Neumann equation (\ref{nle}) for a completely 
positive evolution is a quantum Markovian master equation
(Lindblad equation) \cite{Lind,K1,K2}:
\be \label{8.12} 
\frac{d\rho_t}{dt}=-\frac{i}{\hbar}[H,\rho_t]+
\frac{1}{2\hbar} \sum^{m}_{k=1}\Bigl([ V_{k} \rho_t, V_{k}^{\dagger} ]
+ [ V_{k}, \rho_t V_{k}^{\dagger} ]\Bigr). \ee
This equation in the Liouville space can be written as
\[ \frac{d}{dt}|\rho_t)=\hat \Lambda |\rho_t), \]
where the Liouville superoperator $\hat \Lambda$ is given by
\be \label{sL} \hat \Lambda=-\frac{i}{\hbar}(\hat L_{H}-\hat R_{H})+
\frac{1}{2 \hbar}
\sum^m_{k=1}\Bigl( 2\hat L_{V_{k}} \hat R_{V^{\dagger}_{k}}-
\hat L_{V^{ }_{k}} \hat L_{V^{\dagger}_{k}}-
\hat R_{V^{\dagger}_{k}} \hat R_{V^{ }_{k}} \Bigr). \ee

The basic assumption is that the general form of a bounded
superoperator $\hat \Lambda$, given by the Lindblad theorem, is also
valid for an unbounded  superoperator \cite{AL,SS}. 
Another condition imposed on the operators $H, V_{k}, V_{k}^{\dagger}$ 
is that they are functions of the observables $P$ and $Q$ (with
$[Q,P]=i\hbar I$) of the one-dimensional quantum system. 
Let us consider $V_{k}=a_{k}P+b_{k}Q$,  were $k=1,2$, and $a_{k}$, $b_{k}$
are complex numbers, and the Hamiltonian operator $H$ is
\[ H=\frac{1}{2m} P^2+\frac{m \omega^2}{2} Q^2
+\frac{\mu}{2}(PQ+QP) . \]
Then with the notation \cite{SS}:
\[ d_{qq}=\frac{\hbar}{2}\sum_{k=1,2}{\vert a_{k}\vert}^2, \quad
d_{pp}=\frac{\hbar}{2}\sum_{k=1,2}{\vert b_{k}\vert}^2, \]
\[ d_{pq}=-\frac{\hbar}{2}  Re \Bigl( \sum_{k=1,2}a_{k}^{*}b_{k} \Bigr) , 
\quad \lambda=-Im \Bigl( \sum_{k=1,2}a_{k}^{*}b_{k} \Bigr) \]
equation (\ref{8.12})
in the Liouville space can be written as
\[ \frac{d}{dt}|\rho_t)=
\frac{1}{2m} \hat L^{+}_P \hat L^{-}_P+
\frac{m \omega^2}{2} \hat L^{+}_Q \hat L^{-}_Q -
(\lambda-\mu) \hat L^{-}_P \hat L^{+}_Q |\rho_t)+
(\lambda+\mu) \hat L^{-}_Q \hat L^{+}_P  |\rho_t)+ \]
\be \label{8.16} +d_{pp} \hat L^{-}_Q \hat L^{-}_Q |\rho_t)+
d_{qq} \hat L^{-}_P \hat L^{-}_P|\rho_t)
-2d_{pq} \hat L^{-}_P \hat L^{-}_Q |\rho_t), \ee
where $L^{\pm}$ are the multiplication superoperators defined by
\[ \hat L^{-}_A= \frac{1}{i \hbar} (\hat L_A- \hat R_A), \quad
\hat L^{+}_A=\frac{1}{2}(\hat L_A+ \hat R_A). \]
The properties of these superoperators are considered in the appendix.
Equation (\ref{8.16}) is a superoperator form of the well-known
phenomenological dissipative model \cite{Dek,SS}.


\section{Path integral in the general form}


In the coordinate representation the kernel
\[ \varrho(q,q',t)=(q,q'|\varrho_t) \]
of the density operator $|\varrho_t)$ evolves according to the equation
\[ \varrho(q,q',t)=\int dq_{0} dq'_{0} \
{\cal E}(q,q',q_{0},q'_{0},t,t_{0})
\varrho(q_{0},q'_{0},t_{0}). \]
The function
\be \label{ker} {\cal E}(q,q',q_{0},q'_{0},t,t_{0})=
(q,q'|\hat{\cal E}(t,t_{0})|q_{0},{q'}_{0}) \ee
is a kernel of the linear quantum operation $\hat{\cal E}(t,t_{0})$.
Let the Liouville superoperator $\hat \Lambda_{t}$ be time independent,
i.e. the quantum operation $ \hat{\cal E}(t,t_{0})$ is given by
\be \label{qo3} \hat{\cal E}(t,t_{0})=exp \{ (t-t_{0}) \hat\Lambda \} . \ee

{\bf Proposition 1.}
{\it Let $\{\hat{\cal E}(t,t_{0}),t\ge t_0\}$ be a superoperator
semigroup on operator space $\overline {\cal H}$
\[ \hat{\cal E}(t_0,t_{0})=\hat I, \quad
\hat{\cal E}(t,t_0)=\hat{\cal E}(t,t_{1})\hat{\cal E}(t_1,t_0) , \]
where $t\ge t_1\ge t_0$ such that the infinitesimal generator
$\hat \Lambda$ of this semigroup is defined by (\ref{qo3}).
Then the path integral for kernel (\ref{ker}) of the quantum operation
$\hat{\cal E}(t,t_{0})$ has the following form:
\be \label{pathint} 
{\cal E}(q,q',q_{0},q'_{0},t,t_{0})=
\int {\cal D} q {\cal D} q' {\cal D} p {\cal D} p' \
 exp{ \int^{t}_{t_{0}} dt \Bigl(\frac{\imath}{\hbar}[\dot{q}
p-\dot{q'}p']+ \Lambda_{S}(q,q',p,p') \Bigr) } . \ee
This form is the integral over all trajectories in the
double phase space with the constraints that
$q(t_{0})=q_{0}$, $q(t)=q$, $q'(t_{0})=q'_{0}$, $q'(t)=q'$ and the measure
\[ {\cal D} q=\prod_{t} dq(t) \ , \quad
{\cal D} p=\prod_{t} \frac{dp(t)}{2 \pi \hbar} . \]
The symbol $\Lambda_{S}(q,q',p,p')$ of the Liouville superoperator
$\hat \Lambda$ is connected with the kernel $\Lambda(q,q',y,y')$ by
\[ \Lambda_{S}(q,q',p,p')= \int dy dy' \Lambda(q,q',y,y') \cdot
 exp{-\frac{\imath}{\hbar}[(q-y)p-(q'-y')p']}, \]
where $\Lambda(q,q',y,y')=(q,q'|\hat\Lambda |y,y')$ and
\[ \Lambda(q,q',y,y')= \frac{1}{(2 \pi \hbar)^{2n}}
\int dp dp' \Lambda_{S}(q,q',p,p') \cdot
exp{\frac{\imath}{\hbar}[(q-y)p-(q'-y')p']}. \]
}

{\bf Proof.}

Let us derive path integral form (\ref{pathint})
for quantum operation (\ref{qo3}).

\begin{enumerate}

\item
Let time interval $[t_0,t]$ has $n+1$ equal parts
\[ \tau=\frac{t-t_{0}}{n+1}. \]
Using the superoperator semigroup composition rule
\[ \hat{\cal E}(t,t_{0})=
\hat{\cal E}(t,t_{n}) \hat{\cal E}(t_{n},t_{n-1})...
\hat{\cal E}(t_{1},t_{0}), \]
where $t\ge t_n \ge t_{n-1} \ge ... \ge t_1 \ge t_0$,
we obtain the following integral representation:
\[ {\cal E}(q,q',q_{0},q'_{0},t,t_{0})=
\int dq_{n}dq'_{n} ... dq_{1}dq'_{1} \
{\cal E}(q,q',q_{n},q'_{n},t,t_{n}) \ ...
\ {\cal E}(q_{1},q'_{1},q_{0},q'_{0},t_{1},t_{0}). \]
This representation can be written in the form
\[ {\cal E}(q,q',q_{0},q'_{0},t,t_{0})=
\int \prod^{n}_{k=1} dq_{k}dq'_{k} \prod^{n+1}_{k=1}
{\cal E}(q_{k},q'_{k},q_{k-1},q'_{k-1},t_{k},t_{k-1}). \]
Here $q_{n+1}=q$ and $q'_{n+1}=q'$.

\item
Let us consider the kernel
\[ {\cal E}(q_{k},q'_{k},q_{k-1},q'_{k-1},t_{k},t_{k-1}), \]
of the quantum operation $\hat{\cal E}(t_k,t_{k-1})$.
If the time interval $[t_{k-1},t_{k}]$ is a small,
then in the coordinate representation we have
\[ \varrho(q_{k},q'_{k},t_{k})=
\int dq_{k-1}dq'_{k-1} \
{\cal E}(q_{k},q'_{k},q_{k-1},q'_{k-1},t_{k},t_{k-1})
\varrho(q_{k-1},q'_{k-1},t_{k-1}), \]
and
\[ \varrho (q_{k},q'_{k},t_{k})=(q_{k},q'_{k}|\varrho_{t_k})=
(q_{k},q'_{k}| \hat{\cal E}(t_{k},t_{k-1})| \varrho_{t_{k-1}} )= \]
\[ =\sum^{\infty}_{n=0} \frac{\tau^{n}}{n!}
(q_{k},q'_{k}| \hat \Lambda^{n} |\varrho_{t_{k-1}} )=
(q_{k},q'_{k}|\varrho_{t_{k-1}} )+
(q_{k},q'_{k}| \hat \Lambda |\varrho_{t_{k-1}}) \tau+O(\tau^{2})= \]

\[ =\varrho(q_{k},q'_{k},t_{k-1})+
\tau \int dq_{k-1}dq'_{k-1}
\Lambda(q_{k},q'_{k},q_{k-1},q'_{k-1}) \cdot
\varrho (q_{k-1},q'_{k-1},t_{k-1})+O(\tau^{2})= \]

\[ =\int dq_{k-1} dq'_{k-1} \Bigl(
\delta(q_{k}-q_{k-1}) \delta(q'_{k-1}-q'_{k})+
\tau \Lambda(q_{k},q'_{k},q_{k-1},q'_{k-1})+O(\tau^{2}) \Bigr)
\varrho(q_{k-1},q'_{k-1},t_{k-1}). \]
Therefore, we have
\[ {\cal E}(q_{k},q'_{k},q_{k-1},q'_{k-1},t_{k},t_{k-1})=
\delta(q_{k}-q_{k-1}) \delta(q'_{k-1}-q'_{k})+
\tau \Lambda(q_{k},q'_{k},q_{k-1},q'_{k-1})+O(\tau^{2}). \]

\item
Delta-functions can be written in the form
\[ \delta(q_{k}-q_{k-1}) \delta(q'_{k-1}-q'_{k})=
\int \frac{dp_{k} dp'_{k}}{(2 \pi \hbar)^{2n}} \ 
exp {\frac{i}{\hbar}[(q_{k}-q_{k-1})p_{k}- (q'_{k}-q'_{k-1})p'_{k}]}. \]
Using the relations
\[ (q,q'|p,p') =<q|p><p'|q'>=
\frac{1}{(2 \pi \hbar)^{n}} exp\frac{i}{\hbar}(qp-q'p'), \]
\[ (p,p'|q,q')=<p|q><q'|p'>=
\frac{1}{(2 \pi \hbar)^{n}} exp-\frac{i}{\hbar}(qp-q'p'), \]
we obtain the symbol $\Lambda_{S}(q_{k},q'_{k},p_{k},p'_{k})$
of the Liouville superoperator by
\[ \Lambda(q_{k},q'_{k},q_{k-1},q'_{k-1})=
(q_{k},q'_{k}|\hat\Lambda|q_{k-1},q'_{k-1})= \]
\[ = \int dp_{k} dp'_{k} \ (q_{k},q'_{k}|\hat\Lambda|p_{k},p'_{k})
(p_{k},p'_{k}|q_{k-1},q'_{k-1})=  \]
 \[ = \int dp_{k} dp'_{k}  \ \Lambda_{S}(q_{k},q'_{k},p_{k},p'_{k}) \cdot
(q_{k},q'_{k}|p_{k},p'_{k})(p_{k},p'_{k}|q_{k-1},q'_{k-1})=  \]
\[ =\frac{1}{(2 \pi \hbar)^{2n}} \int dp_{k} dp'_{k}
\ \Lambda_{S}(q_{k},q'_{k},p_{k},p'_{k}) \cdot
exp{ \frac{i}{\hbar}
[(q_{k}-q_{k-1})p_{k}-(q'_{k}-q'_{k-1})p'_{k}] }. \]

\item
The kernel of the quantum operation $\hat{\cal E}(t_{k},t_{k-1})$ is
\[ {\cal E}(q_{k},q'_{k},q_{k-1},q'_{k-1},t_{k},t_{k-1})= \]
\[ =\frac{1}{(2 \pi \hbar)^{2n}} \int dp_{k} dp'_{k} \Bigl( 1+
\tau \Lambda_{S}(q_{k},q'_{k},p_{k},p'_{k})+O(\tau^{2}) \Bigr) \
exp\frac{i}{\hbar}\Bigl((q_{k}-q_{k-1})p_{k}-
(q'_{k}-q'_{k-1})p'_{k}\Bigr). \]

\item
The kernel of the quantum operation $\hat{\cal E}(t,t_{0})$ is
\[ {\cal E}(q,q',q_{0},q'_{0},t,t_{0})= \]
\[ \int \prod^{n}_{k=1} dq_{k} dq'_{k}
\prod^{n+1}_{k=1} \frac{dp_{k} dp'_{k} }{(2 \pi \hbar)^{2n}} \
 exp\frac{i}{\hbar}\sum^{n+1}_{k=1} \Bigl[ (q_{k}-q_{k-1})p_k-
(q'_{k}-q'_{k-1})p'_{k} \Bigr] \cdot \]
\[ \cdot \prod^{n+1}_{k=1}
\Bigl( 1-\tau \Lambda_{S}(q_{k},q'_{k},p_{k},p'_{k})
+O(\tau^{2}) \Bigr). \]

\item
Using
\[ \lim_{n \rightarrow \infty} \prod^{n}_{k=1}(1+\frac{\Lambda_k}{n})=
\lim_{n \rightarrow \infty} \prod^{n}_{k=1}
exp(\frac{\Lambda_k}{n}),  \]
we obtain
\[ {\cal E}(q,q',q_{0},q'_{0},t,t_{0})=
\int \prod^{n}_{k=1} dq_{k} dq'_{k} \prod^{n+1}_{k=1}
\frac{dp_{k} dp'_{k} }{(2 \pi \hbar)^{2n}} \cdot \]
\[ \cdot exp \sum^{n+1}_{k=1} \tau
\Bigl( \frac{i}{\hbar} [\frac{q_{k}-q_{k-1}}{\tau}p_k-
\frac{q'_{k}-q'_{k-1}}{\tau}{p'}_{k} ]+ \
\Lambda_{S}(q_{k},q'_{k},p_{k},p'_{k}) \Bigr). \]

\item
Let $q_{k}$, $q'_{k}$, $p_{k}$, $p'_{k}$ be the values of 
the functions $q(t)$, $q'(t)$, $p(t)$, $p'(t)$ and
$t_{k}=t_{0}+k\tau$, i.e.
\[ q_{k}=q(t_{k}),\ \ q'_{k}=q'(t_{k}), \ \
p_{k}=p(t_{k}) \ , \ \ p'_{k}=p'(t_{k}), \]
where $k=0,1,2,...,n,n+1$. Using
\[ \lim_{\tau \rightarrow 0}\frac{q_{k}-q_{k-1}}{\tau}=\dot q(t_{k-1}), \quad
\lim_{n \rightarrow \infty} \sum^{n+1}_{k=1} A(t_{k})\tau =
\int^{t}_{t_{0}} dt A(t), \]
we obtain the kernel of the quantum operation $\hat{\cal E}(t,t_{0})$
in the path integral form (\ref{pathint}). $\ \ \ \Box$ \\

\end{enumerate}

{\bf Corollary}.
{\it If the dissipative quantum evolution is defined by equation (\ref{neum3}),
then the path integral for the quantum operation kernel has the form
\[ {\cal E}(q,q',q_{0},{q'}_{0},t,t_{0})=\int {\cal D} q {\cal D} p
{\cal D} q' {\cal D} p' \  exp \Bigl( 
{\frac{\imath}{\hbar} ({\cal A}(q,p) - {\cal A}(p',q') )}
+D(q,q',p,p') \Bigr) . \]
Here ${\cal A}(q,p)$ and ${\cal A}(p',q')$ are
action functionals defined by
\be \label{action} 
{\cal A}(q,p)=\int^{t}_{t_{0}} dt \Bigl(\dot{q} p-H(q,p)\Bigr),
\quad {\cal A}(p',q')=\int^{t}_{t_{0}} dt \Bigl(\dot{q'} p'-H(p',q')\Bigr). \ee
The functional $D(q,q',p,p')$ is a time integral of the symbol 
$D_S$ of the superoperator $\hat D$.} \\

The functional $D(q,q',p,p')$ describes the dissipative
part of evolution. \\

{\bf Corollary}.
{\it If the quantum system has no dissipation, i.e.
the quantum system is a closed Hamiltonian system,
then $D(q,q',p,p')=0$ and
the path integral for the quantum operation can be separated
\[ {\cal E}(q,q',q_0,{q'}_0,t,t_0)=
U^{*}(q,q_0,t,t_0)U(q',{q'}_0,t,t_0), \]
where
\[ U(q',q'_0,t,t_0)=\int {\cal D} q' {\cal D} p'
\ exp{-\frac{\imath}{\hbar} {\cal A}(p',q')}, \]
\[ U^{*}(q,q_0,t,t_0)=\int {\cal D} q {\cal D} p
\ exp{\frac{\imath}{\hbar} {\cal A}(p,q)} . \]
}

The path integral for the dissipative quantum systems and
the corresponding quantum operations cannot be separated,
i.e. this path integral is defined in the double phase space.


\section{Path integral for completely positive quantum operation}


Let us consider the Liouville superoperator (\ref{sL})
for the Lindblad equation. \\


{\bf Proposition 2.}
{\it
Let $\{\hat{\cal E}(t,t_{0}),t\ge t_0\}$ be a completely positive
semigroup of linear real trace-preserving superoperators
such that the infinitesimal generators $\hat \Lambda$ of this
semigroup are defined by (\ref{sL}).
Then the path integral for the kernel of the completely 
positive quantum operation $\hat{\cal E}(t,t_{0})$ has the form
\be \label{EL} {\cal E}(q,q',q_{0},{q'}_{0},t,t_{0})=\int {\cal D} q {\cal D} p
{\cal D} q' {\cal D} p' \ {\cal F}(q,q',p,p') \
 exp{\frac{\imath}{\hbar} \Bigl( {\cal A}(q,p) - {\cal A}(p',q') \Bigr)}, \ee
where ${\cal A}(q,p)$ and ${\cal A}(p',q')$ are action functionals (\ref{action})
and the functional ${\cal F}(q,q',p,p')$ is defined as
\be \label{FF} {\cal F}(q,q',p,p')= exp -\frac{1}{2\hbar} \int^{t}_{t_{0}}dt
\sum^m_{k=1} \Bigl( (V^{\dagger}_{k}V_{k})(q,p)+
(V^{\dagger}_{k}V_{k})(p',q')-
2V_{k}(q,p)V^{\dagger}_{k}(p',q') \Bigr). \ee
}

{\bf Proof}.
The kernel of superoperator (\ref{sL}) is
\[ \Lambda(q,q',y,y')=(q,q'_{k}|\hat \Lambda|y,y')=
-\frac{\imath}{\hbar}(<y'|q'><q|H|y>-<q|y><y'|H|q'>)+\]
\[ +\frac{1}{\hbar} \sum^m_{k=1} <q|V_{k}|y><y'|V^{\dagger}_{k}|q'>-
\frac{1}{2\hbar} \sum^m_{k=1}
(<y'|q'><q|V^{\dagger}_{k}V_{k}|y>-
<q|y><y'|V^{\dagger}_{k}V_{k}|q'>) . \]
The symbol $\Lambda_{S}(q,q',p,p')$ of the Liouville superoperator
$\hat \Lambda$ can be derived by
\[ \Lambda(q,q',y,y')=\int dpdp'
\Bigl( -\frac{\imath}{\hbar}( <y'|p'><p'|q'> <q|H_{1}|p><p|y>-
\]
\[ -<q|p><p|y><y'|p'><p'|H_{2}|q'>)+ \]
\[ +\frac{1}{\hbar} \sum^m_{k=1}
<q|V_{k}|p><p|y><y'|p'><p'|V^{\dagger}_{k}|q'> \Bigr) , \]
were the operators $H_1$ and $H_2$ are defined by the relations
\[ H_{1} \equiv H-\frac{\imath}{2}\sum^m_{k=1} V^{\dagger}_{k}V_{k} \ ,
\quad H_{2} \equiv H+\frac{\imath}{2}\sum^m_{k=1} V^{\dagger}_{k}V_{k} . \]
Then the symbol $\Lambda_{S}(q,q',p,p')$
of the Liouville superoperator
(\ref{sL}) can be written in the form
\[ \Lambda_{S}(q,q',p,p')=-\frac{\imath}{\hbar}
\Bigl( H_{1}(q,p)-H_{2}(p',q')+
i\sum^m_{k=1}V_{k}(q,p)V^{\dagger}_{k}(p',q') \Bigr), \]
or
\[ \Lambda_{S}(q,q',p,p')=-\frac{\imath}{\hbar} [ H(q,p)-H(p',q')]-
\]
\[-\frac{1}{2\hbar}\sum^m_{k=1} \Bigl( (V^{\dagger}_{k}V_{k})(q,p)+
(V^{\dagger}_{k}V_{k})(p',q')-
2 V_{k}(q,p)V^{\dagger}_{k}(p',q') \Bigr),  \]
where $H(q,p)$ is a $qp$-symbol of the Hamilton operator $H$
and $H(p,q)$ is a $pq$-symbol of the operator $H$.

In the Hamiltonian case ($V_k=0$), the symbol is given by
\[ \Lambda_{S}(q,q',p,p')=-\frac{\imath}{\hbar}[H(q,p)-H(p',q')]. \]
The path integral for a completely positive quantum operation
kernel has the form
\[ {\cal E}(q,q',q_{0},{q'}_{0},t,t_{0})=\int {\cal D} q {\cal D} p
{\cal D} q' {\cal D} p' \ {\cal F}(q,q',p,p') \
 exp{\frac{\imath}{\hbar} ({\cal A}(q,p) - {\cal A}(p',q') )}. \]
Here ${\cal A}(q,p)$ and ${\cal A}(p',q')$ are
action functionals (\ref{action}), and  the functional
${\cal F}(q,q',p,p')$ is defined by (\ref{FF}). $\ \ \ \Box$ \\

The functional ${\cal F}(q,q',p,p')$ describes the dissipative
part of the evolution and can be called a {\it (double)
phase space influence functional}.
The completely positive quantum operation is described
by the functional (\ref{FF}).\\


{\bf Corollary}.
{\it For the phenomenological dissipative model (\ref{8.16}) the
double phase space path integral has the form (\ref{EL}) with
the functional
\[ {\cal F}(q,q',p,p')= exp \frac{1}{\hbar^2} \int^{t}_{t_{0}} dt
 \Bigl( 2d_{qp} (q-q')(p-p')- \]
\be \label{PM} -d_{qq} (p-p')^2 -d_{pp} (q-q')^2+
i\hbar \lambda (pq'-qp')+i \hbar \mu (q'p'-qp) \Bigr) . \ee
}\\


Using the well-known connection between the phase space (Hamiltonian)
path integral and the configuration space (Lagrangian) path integral 
\cite{SF,Huang}, we can derive the following proposition. \\

{\bf Proposition 3.}
{\it If the symbol $\Lambda_s(q,q',p,p')$ of the Liouville
superoperator can be represented in the form
\be \label{Lam}
\Lambda_{S}(q,q',p,p')=-\frac{\imath}{\hbar}[H(q,p)-H(p',q')]+
D_S(q,q',p,p'), \ee
where
\be \label{Ham} H(q,p)=\frac{1}{2}a^{-1}_{kl}(q)p_kp_l-b_k(q)p_k+c(q) , \ee
\be \label{DS} D_S(q,q',p,p')=-d_k(q,q')p_k+{d'}_k(q,q'){p'}_k+e(q,q'), \ee
then the double phase space path integral (\ref{pathint})
can be represented as a double configuration phase space
path integral
\be \label{FV} {\cal E}(q,q',q_{0},{q'}_{0},t,t_{0})=\int {\cal D} q
{\cal D} q'  \ {\cal F}(q,q') \
exp{\frac{\imath}{\hbar} ({\cal A}(q) - {\cal A}(q') )}, \ee
where 
\be \label{Lag} {\cal A}(q)=\int^t_{t_0} dt \ {\cal L}(q,\dot{q}), \quad 
{\cal L}(q,\dot{q})=\frac{1}{2}a_{kl}(q)\dot{q}_k\dot{q}_l+
a_{kl}(q)b_k(q)\dot{q}_l 
+\frac{1}{2}a_{kl}(q)b_k(q)b_l(q)-c(q) . \ee
This Lagrangian ${\cal L}(q,\dot{q})$ is related to 
the Hamiltonian (\ref{Ham}) by the usual relations
\[ {\cal L}(q,\dot{q})=\dot{q}_kp_k-H(q,p) , \quad
p_k=\frac{\partial {\cal L}}{ \partial \dot{q}} . \]
}

{\bf Proof}.
Substituting (\ref{Lam}) in (\ref{pathint}), we obtain the kernel of
the corresponding quantum operation. Integrating (\ref{pathint})
in $p$ and $p'$, we obtain relation (\ref{FV}) with the functional 
\[ {\cal F}(q,q')=exp\frac{-1}{h^2}\int^t_{t_0} dt \Bigl( 
d_k(q,q')a_{kl}(q) (b_l(q)-\frac{i}{2\hbar}d_l(q,q'))+ \]
\[ +{d'}_k(q,q')a_{kl}(q')(b_l(q')-\frac{i}{2\hbar}{d'}_l(q,q'))+ 
e(q,q')+\delta(0)\Delta(q,q') \Bigr) , \]
where
\[ \Delta(q,q')=-\frac{\hbar^2}{2}\Bigl( ln [ det (a_{kl}(q))]+
ln [ det (a_{kl}(q')) ] \Bigr) . \]
$\ \ \ \Box$ \\

In equation (\ref{FV}) the functional ${\cal F}(q,q')$ can be
considered as the Feynman-Vernon influence functional.
It is known that this functional can be derived by
eliminating the bath degrees of freedom, for example by taking a
partial trace or by integrating them out.
The Feynman-Vernon influence functional describes 
the dissipative dynamics of open systems when we assume 
the von Hove limit for a system-reservoir coupling.
One can describe a quantum system starting
from a unitary evolution of some closed system "system-reservoir" 
if the quantum system is a part of this closed system. 
However, situations can arise where it is difficult or impossible to find a
closed system comprising the given quantum system \cite{FV}-\cite{Str}.

The Feynman path integral is defined for configuration space.
The most general form of quantum mechanical path integral
is defined for the phase space. The Feynman path integral
can be derived from the phase space path integral for the special
form of the Hamiltonian \cite{SF,Huang,LY,IS,R}.
It is known that the path integral for the configuration space
is correct \cite{SF,Huang} only for the Hamiltonian (\ref{Ham}).
The Feynman-Vernon path integral \cite{FV} is defined in the
double configuration space. 
Therefore, this path integral is a special form of 
the double phase space path integral (\ref{pathint}).
The Feynman-Vernon path integral is correct only for the Liouville
superoperator (\ref{Lam}), (\ref{Ham}), (\ref{DS}).
Note that the symbol $\Lambda_s(q,q',p,p')$ for
most of the dissipative and non-Hamiltonian systems
(with completely positive quantum operations)
cannot be represented in the form (\ref{Lam}). \\

{\bf Corollary}.
{\it In the general case, the completely
positive quantum operation cannot be represented
as the double configuration space path integral (\ref{FV}). } \\

For example, the double phase space path integral (\ref{PM})
for the phenomenological dissipative model (\ref{8.16})
has the term $pp'$. Therefore this model and
the Liouville symbol $\Lambda_s(q,q',p,p')$ for this model
cannot be represented in the form (\ref{Lam}), (\ref{Ham}), (\ref{DS}).

\section{On the continuous-variables quantum computation 
by quantum operations with mixed states}

The usual models of a quantum computer deal only with
the discrete variables, unitary quantum operations (gates) and pure states.
Many quantum variables such as position and momentum are continuous.
The use of continuous-variable quantum computing \cite{CV1,CV2,CV3}
allows information to be encoded and processed much more compactly
and efficiently than with discrete variable computing.
Quantum computation using continuous variables is an alternative
approach to quantum computations with discrete variables.

In the models with unitary quantum operations on pure states 
it is difficult or impossible to deal formally 
with measurements, dissipation, decoherence and noise.
It turns out that the restriction to pure states and 
unitary gates is unnecessary \cite{TarJPA}.
In \cite{TarJPA}, a model of quantum computations
by quantum operations with mixed states was constructed.
It is known that the measurement is described by quantum operations. 
The measurement quantum operations are the special case of 
quantum operations on mixed states. 
The von Neumann measurement quantum operation as 
a nonlinear quantum gate is realized in \cite{TarJPA}.
The continuous quantum measurement is described by the path integrals
\cite{Caves,Mensky, Mensky1,Mensky2}. 
Therefore, the path integral for quantum operations can be useful 
for continuous-variables quantum operations on mixed states.
Quantum computation by quantum operations with mixed states is
considered \cite{TarJPA} for discrete variables only. 
Some points of the model of the continuous-variable quantum 
computations with mixed states are considered in this section.
The double phase space path integral can be useful for 
the continuous-variables quantum gates on mixed states.

The main steps of the continuous-variables generalization 
of quantum computations by quantum operations with mixed states
are following.

\begin{enumerate}

\item
The state $|\rho(t))$ of the discrete-variable quantum computation 
with mixed states \cite{TarJPA} is a superposition of basis elements
\begin{equation} \label{rho3}
|\rho(t))=\sum^{N-1}_{\mu=0} |\mu)\rho_{\mu}(t),
\end{equation}
where $\rho_{\mu}(t)=(\mu|\rho(t))$ are real numbers (functions).
The basis $|\mu)$ of the discrete-variable Liouville space 
$\overline{\cal H}^{(n)}$ is defined \cite{TarJPA} by 
\begin{equation} \label{mu}
|\mu)=|\mu_1...\mu_n)=\frac{1}{\sqrt{2^{n}}}|\sigma_{\mu})=
\frac{1}{\sqrt{2^{n}}}|\sigma_{\mu_1}
\otimes  ... \otimes \sigma_{\mu_n}), \end{equation}
where $\sigma_{\mu}$ are Pauli matrices, 
$N=4^{n}$, each $\mu_i \in \{0,1,2,3\}$ and
\begin{equation} \label{mu2}
(\mu|\mu')=\delta_{\mu \mu'} \ , \quad
\sum^{N-1}_{\mu=0} |\mu)(\mu|=\hat I  
\end{equation}
is the discrete-variable computational basis.

The state $|\rho(t))$ of the continuous-variable quantum 
computation at any point of time can be considered as 
a superposition of basis elements
\begin{equation} \label{rho}
|\rho(t))=\int dx \int dx' |x,x')\rho(x,x',t),
\end{equation}
where $\rho(x,x',t)=(x,x'|\rho(t))$ are the density matrix elements.
The basis $|x,x')$ of the continuous-variable operator space 
$\overline{\cal H}$ is defined by $|x,x')=||x><x'|)$, where 
\begin{equation} \label{xx}
(x,x'|y,y')=\delta(x-x')\delta(y-y'), \quad
\int dx \int dx' |x,x')(x,x'|=\hat I  \end{equation}
can be considered as a continuous-variable computational basis.

\item
In the discrete-variable computational basis $|\mu)$ any linear
quantum operation $\hat {\cal E}$ acting on n-qubits mixed (or pure)
states can be represented as a quantum four-valued logic gate \cite{TarJPA}:
$\hat{\cal E}$ on n-ququats can be given by
\begin{equation} \hat{\cal E}=\sum^{N-1}_{\mu=0}\sum^{N-1}_{\nu=0}
{\cal E}_{\mu \nu} \ |\mu)(\nu| , \end{equation}
where $N=4^{n}$,
\begin{equation} {\cal E}_{\mu \nu}=\frac{1}{2^n}
Tr\Bigl(\sigma_{\mu} \hat {\cal E} (\sigma_{\nu}) \Bigr), \end{equation}
and  $\sigma_{\mu} =\sigma_{\mu_1} \otimes ...  \otimes \sigma_{\mu_n}$.

In the continuous-variable computational basis $|x,x')$ any linear
quantum operation $\hat {\cal E}$ acts on mixed (or pure)
states can be represented as a continuous-variable quantum gate:
\begin{equation} 
\hat{\cal E}(t_2,t_1)=\int dx \ dx' \ dy \ dy' \
{\cal E}(x,x',y,y',t_2,t_1) \ |x,x')(y,y'| , \end{equation}
where
${\cal E}(x,x',y,y',t_2,t_1)=(x,x'|\hat {\cal E}(t_2,t_1)|y,y')$
is a kernel of the real trace-preserving positive 
(or completely positive) superoperator $\hat {\cal E}(t_2,t_1)$ 
This quantum operation can be considered as a 
continuous-variable quantum gate.

The continuous quantum measurement which is described by the 
path integral \cite{Caves,Mensky, Mensky1,Mensky2} is the 
special case of the continuous-variable quantum gate.
The path integral for the quantum operations can be useful 
for all continuous-variables quantum operations on mixed states.

\item
Many quantum variables, such as position and momentum are continuous.
The use of continuous-variable quantum computing \cite{CV1,CV2,CV3}
allows information to be encoded and processed much more compactly
and efficiently than with discrete variable computing.

Mixed states subjected to the general quantum operations could 
increase efficiency. This increase is connected with the increasing 
number of computational basis elements for operator Hilbert space.
A pure state of the quantum systems is an element
of functional Hilbert space ${\cal H}$.
A mixed state of the system is an element $|\rho)$ of 
the operator Hilbert space $\overline{\cal H}$.
A mixed state of the system can be considered as an element 
$\rho(x,x',t)$ of the double functional Hilbert space 
${\cal H}\otimes{\cal H}$.  

The use of continuous-variable quantum computation by quantum 
operations with mixed states can increase efficiency 
compared with discrete variable computing.

\end{enumerate}

\section{Conclusion}

The usual quantum computer model deal only with
the discrete variables, unitary quantum operations and pure states.
It is known that many of quantum variables, such as position and 
momentum are continuous.
The use of continuous-variable quantum computing \cite{CV1,CV2,CV3}
allows information to be encoded and processed much more 
efficiently than in discrete-variable quantum computer.
Quantum computation using continuous variables is an alternative
approach to quantum computations with discrete variables.

The quantum computation by quantum operations with mixed states is
considered in \cite{TarJPA}. 
It is known that the measurement is described by quantum operations. 
The measurement quantum operations are the special case of 
quantum operations on mixed states. 
The von Neumann measurement quantum operation as a nonlinear quantum
gate is realized in Ref. \cite{TarJPA}.
The continuous quantum measurement is described by the path integrals
\cite{Caves,Mensky, Mensky1,Mensky2}. 
Therefore, the path integral for quantum operations can be useful 
for continuous-variables quantum operations on mixed states.
Quantum computation by quantum operations with mixed states is
considered \cite{TarJPA} only for discrete variables. 
The model of continuous-variable quantum computations with mixed 
states will be suggested in the next publication.
The double phase space path integral can be useful for 
continuous-variables quantum gates on mixed states.

Let us note the second application of double phase space path integral.
The path integral formulation of the quantum statistical mechanics leads
to the powerful simulation scheme \cite{MP} for the molecular dynamics.
In the past few years the statistical mechanics of non-Hamiltonian
systems was developed for the molecular dynamical simulation purpose
\cite{Tuk1,Tuk2,Ram,Tarmplb,Sergi}.
The suggested path integral can be useful for the application in
the non-Hamiltoniam statistical mechanics of quantum \cite{TN1,TN2} 
and quantum-classical systems \cite{JCP1,JCP2}.


\section*{Acknowledgment}

This work was partially supported by the RFBR grant No. 02-02-16444.

\section*{Appendix}

For the concept of Liouville space and superoperators see
\cite{Cra}-\cite{kn2}.

\subsection*{A.1. Operator space}

The space of linear operators acting on a Hilbert space ${\cal H}$
is a complex linear space $\overline {\cal H}$.
We denote an element $A$ of $\overline{\cal H}$ by a ket-vector $|A)$.
The inner product of two elements $|A)$ and
$|B)$ of $\overline{\cal H}$ is defined as $(A|B)=Tr(A^{\dagger} B)$.
The norm $\|A\|=\sqrt{(A|A)}$ is the Hilbert-Schmidt norm of
operator $A$. A new Hilbert space $\overline{\cal H}$ with the
inner product is called Liouville space attached to ${\cal H}$
or the associated Hilbert space, or Hilbert-Schmidt space
\cite{Cra}-\cite{kn2}.

The X-representation uses eigenfunctions $|x>$ of the operator $X$.
In general, the operator $X$ can be an unbounded operator. 
This operator can have a continuous spectrum. 
This leads us to consider rigged Hilbert space 
\cite{R1,R2,kn1,kn2} (Gelfand triplet) \
${\cal B} \subset {\cal H}={\cal H}^{*} \subset {\cal B}^{*}$
and associated operator space.
The rigged operator Hilbert space can be considered as 
the usual rigged Hilbert space for the operator kernels.

Let the set $\{|x>\}$ satisfy the following conditions:
\[ <x|x'>=\delta(x-x') \ , \quad \int dx |x><x|=I . \]

Then $|x,x')=||x><x'|)$ satisfies
\[ (x,x'|y,y')=\delta(x-x')\delta(y-y'), \quad
\int dx \int dx' |x,x')(x,x'|=\hat I . \]
For an arbitrary element $|A)$ of $\overline{\cal H}$, we have
\begin{equation} \label{|A)}
|A)=\int dx \int dx' |x,x')(x,x'|A) , 
\end{equation}
where $(x,x'|A)$ is a kernel of the operator $A$ such that
\[ (x,x'|A)=Tr( (|x><x'|)^{\dagger}A )=
Tr( |x'><x| A )=<x|A|x'>=A(x,x') . \]
An operator $\rho$ of density matrix
can be considered as an element $|\rho)$ of the Liouville
(Hilbert-Schmidt) space $\overline{\cal H}$.
Using equation (\ref{|A)}), we obtain 
\begin{equation} \label{|rho)} 
|\rho)= \int dx \int dx' |x,x')(x,x'|\rho) \ , 
\end{equation}
where the trace is represented by
\[  Tr\rho=(I|\rho)=\int dx \ (x,x|\rho)=1. \]

\subsection*{A.2. Superoperators}

Operators, which act on $\overline{\cal H}$, are called superoperators
and we denote them in general by the hat.
A superoperator is a map which maps operator to operator.

For an arbitrary superoperator $\hat \Lambda$ on
$\overline{\cal H}$, which is defined by
$\hat \Lambda|A)=|\hat \Lambda(A) )$,
we have
\[ (x,x'|\hat \Lambda|A)=\int dy \int dy'
(x,x'|\hat \Lambda|y,y') (y,y'|A)=
\int dy \int dy'
\Lambda(x,x',y,y') A(y,y') , \]
where $\Lambda(x,x',y,y')=(x,x'|\hat \Lambda|y,y')$
is a kernel of the superoperator $\hat \Lambda$.

Let $A$ be a linear operator in the Hilbert space ${\cal H}$.
We can define the multiplication superoperators
$\hat L_{A}$ and $\hat R_{A}$ by the following equations:
\[ \hat L_{A}|B)=|AB) \ , \quad \hat R_{A}|B)=|BA). \]

The superoperator kernels can be easy derived.
For example, in the basis $|x,x')$ we have
\[ (x,x'|\hat L_{A}|B)=\int dy \int dy'
(x,x'|\hat L_{A}|y,y')(y,y'|B)=
\int dy \int dy' L_{A}(x,x',y,y') B(y,y'). \]
Using
\[ (x,x'|AB)=<x|AB|x'>=
\int dy \int dy' <x|A|y><y|B|y'><y'|x'> , \]
we obtain the kernel of the left multiplication superoperator
\[ L_{A}(x,x',y,y') =<x|A|y><x'|y'>=A(x,y)\delta(x'-y'). \]

A superoperator $\hat{\cal E}^{\dagger}$ is called the adjoint
superoperator for $\hat{\cal E}$ if
$(\hat{\cal E}^{\dagger}(A)|B)=(A|\hat{\cal E}(B))$
for all $|A)$ and $|B)$ from $\overline{\cal H}$.
For example, if $\hat{\cal E}=\hat L_{A}\hat R_{B}$, then
$\hat{\cal E}^{\dagger}=\hat L_{A^{\dagger}}\hat R_{B^{\dagger}}$.
If $\hat{\cal E}=\hat L_{A}$, then
$\hat{\cal E}^{\dagger}=\hat L_{A^{\dagger}}$.

Left superoperators $\hat L^{\pm}_{A}$ are defined
as Lie and Jordan multiplication by the relations
\[ \hat L^{-}_A B=\frac{1}{i\hbar}(AB-BA) , \ \
\hat L^{+}_AB=\frac{1}{2}(AB+BA) . \]
The left superoperator $\hat L^{\pm}_A$ and the
right superoperator $\hat R^{\pm}_A$ are connected by
$\hat L^{-}_A=-\hat R^{-}_A$, \ \ $\hat L^{+}_A=\hat R^{+}_A $.
An algebra of the superoperators $\hat L^{\pm}_{A}$
is defined \cite{Tarmsu} by \\
(1) the Lie relations
\[ \hat L^{-}_{A \cdot B}=\hat L^{-}_{A} \hat L^{-}_{B}-
\hat L^{-}_{B} \hat L^{-}_{A} , \]
(2) the Jordan relations
\[ \hat L^{+}_{(A \circ B)\circ C}+\hat L^{+}_{B}\hat L^{+}_{C}\hat L^{+}_{A}
+\hat L^{+}_{A}\hat L^{+}_{C}\hat L^{+}_{B}=
\hat L^{+}_{A\circ B}\hat L^{+}_{C}+
\hat L^{+}_{B\circ C}\hat L^{+}_{A}+\hat L^{+}_{A\circ C}\hat L^{+}_{B} , \]
\[ \hat L^{+}_{(A\circ B)\circ C}+\hat L^{+}_{B}\hat L^{+}_{C}\hat L^{+}_{A}+
\hat L^{+}_{A}\hat L^{+}_{C}\hat L^{+}_{B}=
\hat L^{+}_{C}\hat L^{+}_{A\circ B}+\hat L^{+}_{B}\hat L^{+}_{A\circ C}+
\hat L^{+}_{A}\hat L^{+}_{B\circ C} , \]
\[ \hat L^{+}_{C}\hat L^{+}_{A\circ B}+\hat L^{+}_{B}\hat L^{+}_{A\circ C}+
\hat L^{+}_{A}\hat L^{+}_{B\circ C}=
\hat L^{+}_{A\circ B}\hat L^{+}_{C}+
\hat L^{+}_{B\circ C}\hat L^{+}_{A}+\hat L^{+}_{A\circ C}\hat L^{+}_{B} , \]
(3) the mixed relations:
\[ \hat L^{+}_{A \cdot B}=
\hat L^{-}_{A}\hat L^{+}_{B}-\hat L^{+}_{B}\hat L^{-}_{A} , \quad
\hat L^{-}_{A \circ B}=\hat L^{+}_{A}\hat L^{-}_{B}+
\hat L^{+}_{B}\hat L^{-}_{A} , \]
\[ \hat L^{+}_{A \circ B}=\hat L^{+}_{A}\hat L^{+}_{B}-\frac{\hbar^{2}}{4}
\hat L^{-}_{B}\hat L^{-}_{A} , \quad
\hat L^{+}_{B}\hat L^{+}_{A}-\hat L^{+}_{A}\hat L^{+}_{B}=
-\frac{\hbar^{2}}{4} \hat L^{-}_{A \cdot B} . \]
Here we use the notations
\[ A \cdot B=\frac{1}{i \hbar}(AB-BA), \ \
 A \circ B=\frac{1}{2}(AB+BA) . \]


\end{document}